\documentstyle[12pt]{article}
\begin{document}\begin{flushright}\thispagestyle{empty}
OUT--4102--82\\
27 September 1999\\
hep-th/9909185                                       \end{flushright}\vfill
                                                     \begin{center}{\Large\bf
Dimensionally continued multi-loop gauge theory      }\vglue 10mm{\large\bf
D.~J.~Broadhurst                                     }\vglue 5 mm{\large
Physics Department, Open University,
Milton Keynes MK7 6AA, UK                          \\[3pt]{\tt
http://physics.open.ac.uk/$\;\widetilde{}$dbroadhu \\[3pt]
D.Broadhurst@open.ac.uk                            }}\end{center}\vfill
                                                   \noindent{\bf Abstract}\quad
  A dimensionally continued background-field method makes the rationality of
the 4-loop quenched QED beta function far more reasonable than had previously
appeared. After 33 years of quest, dating from Rosner's discovery of 3-loop
rationality, one finally sees cancellation of zeta values by the trace
structure of individual diagrams. At 4-loops, diagram-by-diagram cancellation
of $\zeta(5)$ does not even rely on the values of integrals at $d=4$. Rather,
it is a property of the rational functions of $d$ that multiply elements of the
full $d$-dimensional basis. We prove a lemma: the basis consists of slices of
wheels. We explain the previously mysterious suppression of $\pi^4$ in massless
gauge theory. The 4-loop QED result $\beta_4=-46$ is obtained by setting $d=4$
in a precisely defined rational polynomial of $d$, with degree 11. The other 5
rational functions vanish at $d=4$.
\vfill{}\newpage\setcounter{page}{1}
\newcommand{\df}[2]{\mbox{$\frac{#1}{#2}$}}
\newcommand{\ep}{\varepsilon}
\newcommand{\pslash}{p\llap{/\kern-0.7pt}}
\newcommand{\Dslash}{D\llap{/\kern1.5pt}}

\subsection*{1.\quad Introduction}

Connes and Kreimer~\cite{CK}
have recently shown that dimensional regularization
of Feynman diagrams is underwritten by the uniqueness of
the solution of Hilbert's 21st problem~\cite{H21}.
Here, we celebrate this welcome legitimization
of current practice in perturbative quantum field theory,
by finding a rational function of the spacetime dimension $d$
that yields, at $d=4$, the 4-loop term~\cite{4LQ,qbn}
in the beta function of quenched QED. For the first time,
suppression of zeta values is seen diagram by diagram.
We prove the suppression
of $\pi^4$ in the $\overline{\rm MS}$-renormalized
3-loop single-scale Green functions of any massless gauge theory.

There are 6 methods of calculating the 3-loop term~\cite{JLR}
$\beta_3=-2$ in the beta function
\begin{equation}
\beta(a):=\frac{d\log a}{d\log\mu^2}=\sum_{n>0}\beta_n a^n
=\df43a+4a^2-2a^3-46a^4+O(a^5)
\label{beta}
\end{equation}
of quenched QED, with a coupling $a:=\alpha/4\pi$. As described
in~\cite{BDK}, they are as follows.

        M1:\quad Dyson-Schwinger skeleton
                 expansion~\cite{qbn,JLR,BDK,JWB,DRR}.\\
\indent M2:\quad Integration by parts of massive bubble
                 diagrams~\cite{QED,beta}.\\
\indent M3:\quad Integration by parts of massless two-point
                 diagrams~\cite{CT}.\\
\indent M4:\quad Infrared rearrangement of massless bubble
                 diagrams~\cite{4LQ,AAV,IRA,GKL}.\\
\indent M5:\quad Propagation in a background field~\cite{FFS,CGS,GG2}.\\
\indent M6:\quad Crewther connection to deep-inelastic processes~\cite{BK}.

For the 4-loop term, one does not know how
to use M3 directly: there is as yet
no algorithm for 4-loop 2-point functions.
The historical progression through other methods was as follows.
Method M4 first gave $\beta_4=-46$ in~\cite{4LQ}. Then it was noted
in~\cite{BK} that the result was consistent with deep-inelastic
results~\cite{DIS}
by virtue of the exactness of the Crewther~\cite{RJC} connection
M6 in the quenched abelian case. Progress with 4-loop massive
bubbles in M2 led to the 4-loop beta function of a general~\cite{beta}
gauge theory, confirming the particular case of QED. Recently~\cite{qbn}
we used the Dyson-Schwinger method M1, which was shown to be very
efficient. However, in none of these 4 analyses does one gain
an understanding of how the rationality of $\beta_4=-46$ comes about;
all 4 involve intricate cancellations of zeta values
between diagrams with quite different momentum flows.

The attentive reader will have noticed that one stone lay unturned:
the background-field method M5. Very recently we attempted it,
and found it to be wonderfully user-friendly: 8 lines of code suffice.
This is because it gives $\beta_4$ directly in terms of 8 three-loop diagrams,
whereas in~\cite{4LQ,qbn} a reduction to 3 loops was achieved indirectly,
via nullification of four-loop diagrams, which are much more numerous.
Moreover, we find that in a background field the cancellation of zeta values
is {\em much} less obscure.

Our method follows immediately from the telling observation~\cite{JB} that
$d(\beta(a)/a)/d a=\sum_{n>1} (n-1)\beta_n a^{n-2}$
is given by the radiative corrections to the photon
self-energy of massless quenched QED,
in a background field. Consider the momentum-space correlator
\begin{equation}
i\int dx~e^{i k\cdot x}~\langle0|T(J_\mu(x)J_\nu(0))|0\rangle
=(k_\mu k_\nu-k^2g_{\mu\nu})\left\{\Pi_0(k^2)+\Pi_2(k^2)F^2
+O(F^4)\right\}
\label{bf}
\end{equation}
of the electromagnetic current $J_\mu:=\overline\psi\gamma_\mu\psi$
in a background electromagnetic
field $F_{\mu\nu}$, with $F^2:=\langle F_{\mu\nu}F^{\mu\nu}\rangle$.
After struggling~\cite{qbn} to decode the rationality of $\beta_4$
via the 4-loop contribution to $\Pi_0$, I judged it more prudent to tackle
the finite 3-loop contribution in
\begin{equation}
\Pi_2(k^2)=\frac{\beta_2a}{6k^4}\left\{
1+\left(\frac{2\beta_3}{\beta_2}\right){a}
+\left(\frac{3\beta_4}{\beta_2}\right){a}^2
+O({a}^3)\right\}
\label{Pi2}
\end{equation}

\subsection*{2.\quad Three-loop beta function}

First we dimensionally continue Rosner's result, $2\beta_3/\beta_2=-1$.
Consider the $O(F^2)$ term of~(\ref{bf}) in $d:=4-2\ep$ dimensions,
with a dimensionless coupling $\overline{a}:=(4\pi)^{\ep}g(k^2)a$, where
\begin{equation}
g(k^2):=\frac{\Gamma(1+\ep)\Gamma^2(1-\ep)}{k^{2\ep}\Gamma(1-2\ep)}
\label{ab}
\end{equation}
absorbs the $\Gamma$ functions and momentum
dependence of one-loop integration.
Let the $d$-dimensional form of~(\ref{Pi2}) be written as
\begin{equation}
\overline{\Pi}_2(k^2)=\frac{\overline\beta_2\overline{a}}{2(d-1)k^4}
\left\{1
+\left(\frac{2\overline\beta_3}{\overline\beta_2}\right)\overline{a}
+\left(\frac{3\overline\beta_4}{\overline\beta_2}\right)\overline{a}^2
+O(\overline{a}^3)\right\}
\label{Pi2d}
\end{equation}
with bars denoting analytic continuation in $d$. Then the one-loop
self-energy yields
\begin{equation}
\overline\beta_2=\frac{(6-d)(d-2)}{d}{\rm Tr}(1)
\label{b2}
\end{equation}
where Tr$(1)$ is whatever one chooses to take for the trace
of the unit matrix in the Clifford algebra of $d$ dimensions.
At $d=4$, where certainly Tr$(1)=4$, we get $\beta_2=4$.

Now consider the $d$-dimensional analysis
for $\overline\beta_3$. Integration by parts~\cite{CT} gives
the first radiative correction in~(\ref{Pi2d})
in terms of $\Gamma$ functions:
\begin{equation}
\frac{2\overline\beta_3}
{\overline\beta_2}=R_{2,0}+R_{2,3}\overline\zeta_3
\label{b3}
\end{equation}
where $R_{2,0}$ and $R_{2,3}$ are rational functions of $d$ and
\begin{equation}
\overline\zeta_{3}:=
\frac{(d-3)k^{4}}{6(g(k^2)\pi^{d/2})^2}
\int\frac{dP_1dP_2dP_3~\delta(k-P_1)}{P_1^2P_2^2P_3^2
(P_1-P_2)^2(P_2-P_3)^2(P_3-P_1)^2}
\label{w3}
\end{equation}
is a slice, via a $\delta$ function, of the wheel with three spokes:
the tetrahedron. It is a template for
our later construction of a 3-loop basis. At 2 loops, one easily~\cite{CT}
evaluates
\begin{equation}
\overline\zeta_3
=\frac{1}{6\ep^3}\left(
1-\frac{\sec(\pi\ep)\Gamma(1-2\ep)}{\Gamma(1+\ep)\Gamma(1-3\ep)}\right)
=\zeta(3)+\df32\zeta(4)\ep+7\zeta(5)\ep^2+O(\ep^3)
\label{z3}
\end{equation}
which is the sole origin of $\zeta(3):=\sum_{n>0}1/n^3$
in 2-loop 2-point functions at $d=4$.
Exact $d$-dimensional analysis gives the rational functions
\begin{eqnarray}
R_{2,0}&=&-4\frac
{d^6-37d^5+554d^4-4280d^3+17826d^2-37728d+31608}
{(d-1)(d-3)^2(d-6)^3(d-8)}
\label{R20}\\
R_{2,3}&=&6\frac
{(d-2)(d-4)(d-5)(d^4-25d^3+230d^2-920d+1376)}
{(d-1)(d-6)^3(d-8)}
\label{R23}
\end{eqnarray}
This massless abelian analysis is far
simpler than our massive non-abelian work~\cite{GG2,BFT}.
The rationality of $\beta_3=-2$ is seen in
the vanishing of~(\ref{R23}) at $d=4$; the precise
value comes from setting $d=4$ in~(\ref{R20}),
which gives $2\beta_3/\beta_2=-1$.
Before proceeding to $\beta_4$, we prove a lemma,
which lies at the heart of the method and has wider import.

\subsection*{3.\quad Suppression of $\pi^4$
in massless 3-loop gauge theory}

It is well understood~\cite{GPX} why $\zeta(3)$ is the first
irrational number to appear in single-scale processes,
renormalized in the $\overline{\rm MS}$ scheme.
The suppression of a single power of $\pi^2$ is a property of any
massless Lorentz-covariant theory: every bare 2-loop 2-point diagram, with
spacelike momentum $k$, gives
a result of the form $(R_{2,0}+R_{2,3}\overline\zeta_3)\overline{a}^2$,
where $R_{2,0}$ may have $1/\ep^2$ and $1/\ep$ singularities,
while $R_{2,3}$ is regular at $d=4$.
In heavy-quark effective theory,
where the worldline of the heavy quark breaks Lorentz symmetry,
one finds 2-loop combinations of $\Gamma$ functions
that lack~\cite{EFT} the $\pi^2$ suppression of~(\ref{z3}).

It is the experience of several workers that if -- and usually not until --
one {\em combines} the 3-loop diagrams specified by a massless gauge theory,
$\pi^4$ vanishes as well~\cite{GKL,BK,DIS,BKT}.
Multi-loop
colleagues have asked me why this happens. The answer
lies in the output format of the program~{\sc slicer}~\cite{BKT},
which gives the results of subsequent sections.
Following the example of~(\ref{w3}), we define 4 three-loop integrals
that depend only on $d$:
\begin{equation}
\overline\zeta_{5,S}:=
\frac{(d-3)k^{4}}{20(g(k^2)\pi^{d/2})^3}
\int\frac{dP_1dP_2dP_3dP_4~\delta(k-P_S)}{P_1^2P_2^2P_3^2P_4^2
(P_1-P_2)^2(P_2-P_3)^2(P_3-P_4)^2(P_4-P_1)^2}
\label{w4}
\end{equation}
with 4 distinct slices of the 4-spoke wheel, encoded by
$S=L,M,N,Q$, corresponding to
$P_L:=P_1-P_3$, $P_M:=P_1$, $P_N:=P_1-P_2+P_3-P_4$, $P_Q:=P_1-P_2$,
for the external momentum, $k$. Then each integral evaluates
to $\zeta(5)$ at $d=4$, since it comes from slicing the wheel with 4
spokes and we have proved~\cite{wheel} that
the wheel with $n+1$ spokes gives ${2n\choose n}\zeta(2n-1)$, for $n>1$.
The coding was suggested by letters in~\cite{CT}:
the L (ladder) slice is at the 4-point hub, in the $s$-channel;
M (Mercedes-Benz) is at the rim;
N (non-planar) is at the hub, in the $u$-channel;
Q (?) is at a spoke. Remarkably, this {\em completes} a basis.
\begin{quotation}
\noindent{\bf Lemma:}\quad
A bare 3-loop 2-point diagram with external momentum $k$ gives
\begin{equation}
\left\{R_{3,0}+R_{3,3}\overline\zeta_3
+\sum_{S=L,M,N,Q}R_{3,S}\overline\zeta_{5,S}\right\}\overline{a}^3
\label{lemma}
\end{equation}
with coefficients $\{R_{3,S}\mid S=0,3,L,M,N,Q\}$
that are rational functions of $d$.
\end{quotation}
{\em Proof:} This is an extension of the fine analysis
of Chetyrkin and Tkachov~\cite{CT}.
Consider the 5 one-particle-irreducible 4-loop bubble diagrams of Fig.~1.
Generate all tadpole-free trivalent 3-loop 2-point diagrams by slicing these,
to obtain the 12 distinct cases of Fig~2, where each slice is marked.
Whatever the numerators, and whatever the integer powers of the
propagators, integration by parts~\cite{CT} allows one to eliminate
all and only those diagrams that contain at least one
unsliced triangle. Thus we may eliminate 5 of the 12 cases, namely
C2, C3, C4, D1, D2. The final observation is that B3,
the sole case with a sliced chord in Fig.~2,
is rationally related to B2.
The basic one-loop integral in $d$ dimensions is~\cite{GPX}
\begin{equation}
G(a,b):=
\frac{\Gamma(a+b-d/2)}{\Gamma(a)\Gamma(b)}
\frac{\Gamma(d/2-a)\Gamma(d/2-b)}{\Gamma(d-a-b)}
\label{gab}
\end{equation}
with propagators raised to powers $a$ and $b$. Let $\Phi({\rm X})$
be the representative of class X in $d$-dimensional $\phi^3$ theory. Using
only $\Gamma(z+1)=z\Gamma(z)$, we obtain
\begin{equation}
\frac{\Phi({\rm B3})}{\Phi({\rm B2})}=
\frac{G(1,1)G(1+\ep,1+\ep)}{G(1,1+\ep)G(1,1+2\ep)}
=\frac{3d-10}{d-3}
\label{b23}
\end{equation}
Then~(\ref{lemma}) is a basis for the 6 remaining cases,
since~(\ref{w3},\ref{w4}) are independent. $\Box$.

Two comments are in order. First,
the factor $d-3$ in~(\ref{w4}) ensures
that the $\ep$ expansions of $\overline\zeta_{5,S}$ are all pure: at
order $\ep^n$ one encounters {\em multiple}~\cite{BGK} zeta values,
exclusively of weight $n+5$. All mixing derives from the rational
coefficients, of which only $R_{3,0}$ and $R_{3,3}$ may be singular at $d=4$.
Secondly, there are two~\cite{CT} combinations of
$\{\overline\zeta_{5,S}\mid S=L,M,Q\}$
that may be reduced to $\Gamma$ functions. No purpose is served
by making such reductions: one would need 5th order Taylor series for
those $\Gamma$ functions to arrive back at a result for $d=4$ that is
immediately available from~(\ref{lemma}).
Moreover, such reduction would obscure our physical conclusion.
\begin{quotation}
\noindent{\bf Corollary:}\quad
In massless gauge theory,
$\overline{\rm MS}$-renormalized 3-loop single-scale Green functions
do not involve $\pi^4$.
\end{quotation}
{\em Proof:} The lemma shows that $\zeta(4)=\pi^4/90$ may arise
only from~(\ref{z3}).
This would occur only if $R_{3,3}$ is singular. But such a singularity,
in the sum of diagrams, would require a counterterm $\zeta(3)/\ep$
in the coupling. It is proven that
the 3-loop $\beta$ function of any gauge theory is rational~\cite{TVZ},
hence {\em no} such counterterm is available. $\Box$

Now one sees why cancellations of
$\pi^4$ were regarded as surprising
in computations such as~\cite{GKL,DIS}.
The bare $\zeta(4)$ terms generated by~{\sc mincer}~\cite{mincer}
appeared to have several sources: Taylor expansions of
$\Gamma$ functions from diagrams in classes A1, A2, B1, B2, B3.
By contrast, {\sc slicer}~\cite{BKT}
outputs the 6 exact $d$-dimensional rational functions
of the lemma. In our approach, one is bound to
find that $R_{3,3}$ is regular in the sum of diagrams
of a massless gauge theory; {\sc slicer} encodes the suppression
of $\pi^4$, {\em ab initio}.

Note, however, that the 3-loop anomalous mass dimension~\cite{mass3}
involves $\zeta(3)$.
Thus if one expands in the fermion mass, $\pi^4$ may emerge.
Indeed, one finds $\zeta(4)$ in non-abelian terms of the
4-loop anomalous mass dimension~\cite{Kostja,mass}.
Its absence from the 4-loop beta function
is guaranteed by the lemma and serves as a check of the full
result~\cite{beta}.

\subsection*{4.\quad Four-loop beta function}

For $\beta_4$, we need the finite 3-loop term in~(\ref{Pi2}).
Instead of the $7\times5\times3=105$ four-loop diagrams
of~\cite{4LQ}, there are only $5\times3=15$ three-loop diagrams.
These are reduced to 8, by symmetries.
Each  contains a single fermion loop: a hamiltonian circuit with
6 vertices. Consider a bare fermion propagator, with momentum $p$.
Without a background field, it would simply be
$S(p)=i/\pslash$. To take account of the abelian background field,
$F_{\mu\nu}$, one replaces this by
\begin{equation}
S(p+i D)=S(p)-S(p)\Dslash S(p)+S(p)\Dslash S(p)\Dslash S(p)+O(D^3)
\label{SP}
\end{equation}
where $D_\mu$ is an operator in a Fock-Feynman-Schwinger
formalism~\cite{FFS} that is oblivious to the momentum integrations,
feeling only the order of $\gamma$ matrices.
All one needs to know is that $[D_\mu,D_\nu]=-i e F_{\mu\nu}$,
since $D_\mu$ acts as a gauge-covariant derivative
in the external configuration space.
For a modern review of the non-abelian case, see~\cite{AGG}.
Here, we need only obtain the coefficient of
$F^2:=\langle F_{\mu\nu}F^{\mu\nu}\rangle$, using
\begin{equation}
\langle D_\alpha D_\beta D_\mu D_\nu \rangle=
\left(g_{\alpha\nu}g_{\beta\mu}-g_{\alpha\beta}g_{\mu\nu}\right)
\frac{e^2F^2}{2d(d-1)}
\label{abmn}
\end{equation}
for the expectation value of 4 external derivations.
This is contracted with all ways of making 4 ordered insertions
of gamma matrices in the 6 fermion propagators on a hamiltonian circuit
that begins and ends at an external vertex of the 3-loop Feynman diagram.
The tensor in~(\ref{abmn}) implies
that the second-order expansion~(\ref{SP})
suffices for any single propagator: insertion of more than
2 gamma matrices in any one propagator leads to a vanishing contraction.
Thus no infrared divergence is produced.
Simple counting shows that each diagram produces 180 traces,
containing 20 gamma matrices.

This astoundingly user-friendly method is CPU-intensive.
In the Dyson-Schwinger
method~\cite{qbn}, it was possible
to take traces before double differentiation
w.r.t.\ an external photon momentum,
thus limiting the traces to 16 gamma matrices. Moreover, for
the most demanding integrals one needed only traces taken at $d=4$.
Now, the exact handling of two non-commutative algebras
-- Dirac's gamma matrices and Schwinger's covariant derivatives --
requires more machine time, while the physicist relaxes.
Within hours of reading~\cite{CK},
I had typed 8 lines into~{\sc slicer}, one for each of the 8 diagrams,
added a short procedure to generate the $8\times180=1440$ traces,
and farmed the problem out, using a cluster of DecAlpha machines.
Since {\sc slicer}, using {\sc reduce}, is wildly uncompetitive
with Jos Vermaseren's lightning-fast implementation of~{\sc mincer}
in~{\sc form}, it was 2 days later when the answer $-46$ appeared on screen.
However, the object was not speed. Rather it was a better understanding
of cancellation of zeta values, which was mysterious
in the faster Dyson-Schwinger method~\cite{qbn}.

\subsection*{5.\quad Fock-Feynman-Schwinger anatomy}

The Dyson-Schwinger anatomy of $\beta_3$ exhibits cancellation of
$\zeta(3)$ between diagrams in Landau gauge~\cite{JLR},
in Feynman gauge~\cite{qbn,JB}, and indeed in any gauge~\cite{BDK}.
In the background-field method, $\beta_3$ comes from two 2-loop
diagrams, each of which is free of $\zeta(3)$, as $d\to4$,
in any combination of internal and external gauges.
We used a gauge $(q^2g_{\mu\nu}+(\xi_{\rm int}-1)q_\mu q_\nu)/q^4$,
for the internal photon propagator, and contracted~(\ref{bf}) with
$k^2g_{\mu\nu}+(\xi_{\rm ext}-1)k_\mu k_\nu$, where $k$ is the external photon
momentum.
As $\ep\to0$, we found
\begin{eqnarray}
C_3({\rm PE})&=&-\df12
+\xi_{\rm int}\left(\frac{2}{\ep}+1\right)
-\xi_{\rm ext}\left(\frac{1}{\ep}+2\right)
-\xi_{\rm int} \xi_{\rm ext}
\\
C_3({\rm DF})&=&-\df12
-\xi_{\rm int}\left(\frac{2}{\ep}+1\right)
+\xi_{\rm ext}\left(\frac{1}{\ep}+2\right)
+\xi_{\rm int} \xi_{\rm ext}
\\
2\beta_3/\beta_2&=&-1
\end{eqnarray}
where PE and DF identify the photon-exchange and dressed-fermion
contributions. There is no sign of $\zeta(3)$ in either diagram.

This was no accident, as witnessed
by the 4-loop result, obtained in Feynman gauge, for the sake of economy.
As $\ep\to0$, the contributions and total for $3\beta_4/\beta_2$ are
\begin{eqnarray}
C_4({\rm A1})&=&
 \frac{2}{3\ep^2}
-\frac{7}{3\ep}
-\frac{58}{3}
\label{n0}\\
C_4({\rm B2})&=&
 \frac{2}{3\ep^2}
-\frac{4}{3\ep}
+\frac{5}{3}
\\
C_4({\rm B3})&=&
-\frac{3}{2\ep}
-\frac{21}{4}
\\
C_4({\rm C2})&=&
-64\zeta(3)
-\frac{4}{3\ep^2}
+\frac{26}{3\ep}
+79
\label{n1}\\
C_4({\rm C4})&=&
-\frac{32}{3}\zeta(3)
-\frac{4}{3\ep^2}
+\frac{6}{\ep}
+\frac{209}{9}
\\
C_4({\rm D1})&=&
 \frac{320}{3}\zeta(3)
+\frac{4}{3\ep^2}
-\frac{9}{\ep}
-\frac{2695}{18}
\\
C_4({\rm D2})&=&
 64\zeta(3)
-\frac{9}{2\ep}
-\frac{1177}{12}
\\
C_4({\rm E1})&=&
-96\zeta(3)
+\frac{4}{\ep}
+134
\label{n2}\\
3\beta_4/\beta_2&=&-\frac{69}{2}
\end{eqnarray}
with diagrams identified by Fig.~2. There is no sign of $\zeta(5)$
in any diagram.
This, again, is radically different from the Dyson-Schwinger method.

Even more impressive is the distribution of $\zeta(5)$
between the 4 distinct slices in~(\ref{w4}). By construction,
{\sc slicer} is oblivious to the contingency that these integrals
happen to be equal at $d=4$. Thus it records that each is cancelled
separately, in each diagram, by the Dirac-Schwinger
traces that produce 4 exact rational functions of $d$, each vanishing at $d=4$.
Thus, in the background-field method, the absence of $\zeta(5)$
relies neither on conspiracies between different diagrams
nor on any happenstance of 4-dimensional analysis.
This is a reasonable type of rationality, that one may realistically
hope to understand better, by essentially combinatoric methods.

It will be noted that the subleading zeta value,
$\zeta(3)$, is seen in results~(\ref{n1}--\ref{n2}),
for individual diagrams that are slices of the C, D, E topologies of Fig.~1.
However, it should also be noted that this subleading irrational
occurs only at subleading order in $\ep$. Topologies
C and D produce $\zeta(3)/\ep$ singularities, at the level
of individual Dirac-Schwinger traces. These conspicuously
cancel, diagram by diagram.
Thus background-field trace algebra accounts for
6 of the 7 features of rationality up to 4 loops:
the cancellation of $\zeta(3)$ at 3 loops;
of all 4 species of $\zeta(5)$, separately, at 4-loops;
of $\zeta(3)/\ep$ at 4 loops.
Only the final cancellation -- subleading
irrational, subleading in $\ep$ -- entails
conspiracy between Feynman-gauge diagrams. A {\sc mincer} analysis
for all $\xi_{\rm int}$ and $\xi_{\rm ext}$ would be fascinating.

\subsection*{6.\quad Exercise and conclusion}

We have completed an instructive exercise in
dimensional continuation of a {\em finite} quantity.
\begin{enumerate}
\item Evaluate the exact
$d$-dimensional 3-loop radiative correction in~(\ref{Pi2d}).
By the lemma of section~3, this amounts to finding
6 precisely defined rational functions of $d$ in
\begin{equation}
\frac{3\overline\beta_4}{\overline\beta_2}
=R_{3,0}+R_{3,3}\overline\zeta_3
+\sum_{S=L,M,N,Q}R_{3,S}\overline\zeta_{5,S}
\end{equation}
with wheel-slice basis~(\ref{w3},\ref{w4}).
Forget $\pi^4$; in massless gauge theory it never happens.
\item
Verify that $\{R_{3,S}\mid S\neq0\}$ contain the factor $d-4$.
Dissect this, diagram by diagram.
Conclusion: rationality appears
far more reasonable than in any other method.
\item Find the numerator and denominator of $R_{3,0}=N(d)/D(d)$. Solution:
\begin{eqnarray}
N(d)&=&
1215d^{11}
-53433d^{10}
+1072059d^9
-12995191d^8
+105924166d^7
\nonumber\\&&{}
-609433848d^6
+2520429944d^5
-7469717936d^4
+15495188128d^3
\nonumber\\&&{}
-21364053504d^2
+17580978560d
-6532684800
\label{P46}\\
D(d)&=&2(d-1)(d-3)^3(d-5)^2(d-6)^3(d-8)(3d-8)(3d-10)^2
\end{eqnarray}
Hence, in 4-loop quenched QED, $\beta_4=N(4)/2^8 3^2$.
Check that the answer is $-46$.
\end{enumerate}

I conclude that the key to Jonathan Rosner's fine puzzle~\cite{JLR} was
given by Marshall Baker and Kenneth Johnson, in Eq~(3.3) of~\cite{JB}.
Noting the profound work of Alain Connes and Dirk Kreimer~\cite{CK},
one arrives at the nub of the rationality of quenched QED:
dimensional continuation of the {\em derivative} of the scheme-independent
single-fermion-loop Gell-Mann--Low function, via the Fock--Feynman--Schwinger
formalism~(\ref{SP},\ref{abmn}). It remains to be seen whether this can
tell us what comes after $\beta_4=-46$. Hope has risen.

\newpage
\noindent{\bf Acknowledgements:}
Encouragement from Marshall Baker and Jonathan Rosner kept
me believing that there was a key to be found.
Recollection of enjoyable work with
Sotos Generalis and Andrey Grozin
reminded me of~(\ref{abmn}).
Jos Vermaseren encouraged inclusion of my
explanation of the suppression of $\pi^4$.
Alain Connes and Dirk Kreimer
provided the vital impetus for
an exact $d$-dimensional analysis,
by showing me an early version of~\cite{CK}.

\raggedright

\newpage
\vfill
\setlength{\unitlength}{0.014cm}
\newbox\shell
\newcommand{\lbl}[3]{\put(#1,#2){\makebox(0,0)[b]{$#3$}}}
\newcommand{\dia}[1]{\setbox\shell=\hbox{
\begin{picture}(180,200)(-90,-100)#1\end{picture}}\dimen0=\ht
\shell\multiply\dimen0by7\divide\dimen0by16\raise-\dimen0\box\shell}
\newcommand{\blob}{\circle*{10}}
\begin{center}{\bf Fig~1:}\quad The 5 trivalent four-loop bubble
diagrams, presented as chord diagrams\end{center}
\hspace*{2mm}
\dia{
\put(0,0){\circle{100}}
\put(-48,-10){\line(+2,+3){39}}
\put(+48,-10){\line(-2,+3){39}}
\put(-40,-30){\line(+1, 0){80}}
\lbl{0}{-110}{{\bf A}}}
\hspace{2mm}
\dia{
\put(0,0){\circle{100}}
\put(0,-50){\line(0,+1){100}}
\put(-30,-40){\line(0,+1){80}}
\put(+30,-40){\line(0,+1){80}}
\lbl{0}{-110}{{\bf B}}}
\hspace{2mm}
\dia{
\put(0,0){\circle{100}}
\put(-48,-10){\line(+2,+1){86}}
\put(+48,-10){\line(-2,+1){86}}
\put(-40,-30){\line(+1, 0){80}}
\lbl{0}{-110}{{\bf C}}}
\hspace{2mm}
\dia{
\put(0,0){\circle{100}}
\put(-50,0){\line(+1,0){100}}
\put(-30,-40){\line(0,+1){80}}
\put(+30,-40){\line(0,+1){80}}
\lbl{0}{-110}{{\bf D}}}
\hspace{2mm}
\dia{
\put(0,0){\circle{100}}
\put(-50,0){\line(1,0){100}}
\put(-30,-40){\line(+3,+4){60}}
\put(-30,+40){\line(+3,-4){60}}
\lbl{0}{-110}{{\bf E}}}

\vfill

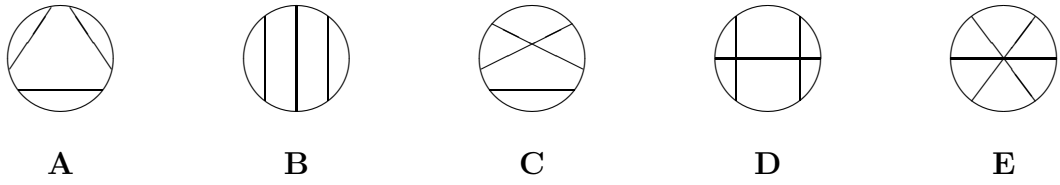
\begin{center}{\bf Fig~2:}\quad The 12 trivalent 3-loop 2-point
diagrams, from slices of Fig.~1\end{center}
\hspace*{10mm}
\dia{
\put(0,0){\circle{100}}
\put(0,-50){\blob}
\put(-48,-10){\line(+2,+3){39}}
\put(+48,-10){\line(-2,+3){39}}
\put(-40,-30){\line(+1, 0){80}}
\lbl{0}{-110}{{\bf A1}}}
\hspace{5mm}
\dia{
\put(0,0){\circle{100}}
\put(0,50){\blob}
\put(-48,-10){\line(+2,+3){39}}
\put(+48,-10){\line(-2,+3){39}}
\put(-40,-30){\line(+1, 0){80}}
\lbl{0}{-110}{{\bf A2}}}
\hspace{5mm}
\dia{
\put(0,0){\circle{100}}
\put(-20,45){\blob}
\put(0,-50){\line(0,+1){100}}
\put(-30,-40){\line(0,+1){80}}
\put(+30,-40){\line(0,+1){80}}
\lbl{0}{-110}{{\bf B1}}}
\hspace{5mm}
\dia{
\put(0,0){\circle{100}}
\put(-50,0){\blob}
\put(0,-50){\line(0,+1){100}}
\put(-30,-40){\line(0,+1){80}}
\put(+30,-40){\line(0,+1){80}}
\lbl{0}{-110}{{\bf B2}}}
\\\hspace*{10mm}
\dia{
\put(0,0){\circle{100}}
\put(0,0){\blob}
\put(0,-50){\line(0,+1){100}}
\put(-30,-40){\line(0,+1){80}}
\put(+30,-40){\line(0,+1){80}}
\lbl{0}{-110}{{\bf B3}}}
\hspace{5mm}
\dia{
\put(0,0){\circle{100}}
\put(0,50){\blob}
\put(-48,-10){\line(+2,+1){86}}
\put(+48,-10){\line(-2,+1){86}}
\put(-40,-30){\line(+1, 0){80}}
\lbl{0}{-110}{{\bf C1}}}
\hspace{5mm}
\dia{
\put(0,0){\circle{100}}
\put(-48,10){\blob}
\put(-48,-10){\line(+2,+1){86}}
\put(+48,-10){\line(-2,+1){86}}
\put(-40,-30){\line(+1, 0){80}}
\lbl{0}{-110}{{\bf C2}}}
\hspace{5mm}
\dia{
\put(0,0){\circle{100}}
\put(-45,-20){\blob}
\put(-48,-10){\line(+2,+1){86}}
\put(+48,-10){\line(-2,+1){86}}
\put(-40,-30){\line(+1, 0){80}}
\lbl{0}{-110}{{\bf C3}}}
\\\hspace*{10mm}
\dia{
\put(0,0){\circle{100}}
\put(0,-50){\blob}
\put(-48,-10){\line(+2,+1){86}}
\put(+48,-10){\line(-2,+1){86}}
\put(-40,-30){\line(+1, 0){80}}
\lbl{0}{-110}{{\bf C4}}}
\hspace{5mm}
\dia{
\put(0,0){\circle{100}}
\put(-45,20){\blob}
\put(-50,0){\line(+1,0){100}}
\put(-30,-40){\line(0,+1){80}}
\put(+30,-40){\line(0,+1){80}}
\lbl{0}{-110}{{\bf D1}}}
\hspace{5mm}
\dia{
\put(0,0){\circle{100}}
\put(0,50){\blob}
\put(-50,0){\line(+1,0){100}}
\put(-30,-40){\line(0,+1){80}}
\put(+30,-40){\line(0,+1){80}}
\lbl{0}{-110}{{\bf D2}}}
\hspace{5mm}
\dia{
\put(0,0){\circle{100}}
\put(0,50){\blob}
\put(-50,0){\line(1,0){100}}
\put(-30,-40){\line(+3,+4){60}}
\put(-30,+40){\line(+3,-4){60}}
\lbl{0}{-110}{{\bf E1}}}

\vfill{}

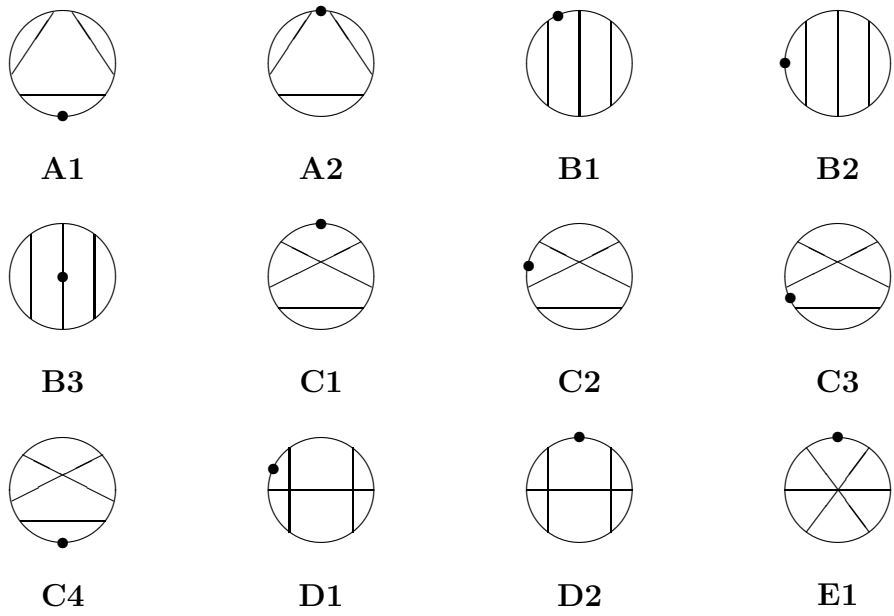
\end{document}